\documentstyle[sprocl,epsfig]{article}

\def\be{\begin{equation}}
\def\ee{\end{equation}}
\def\bea{\begin{eqnarray}}
\def\eea{\end{eqnarray}}

\begin{document}

\title{FIRST PRINCIPLE QUADRATIC AND LINEAR MASS OPERATORS,
 QUARKONIUM SPECTRUM AND \\ REGGE TRAJECTORIES}

\author{M. BALDICCHI}

\address{Dipartimento di Fisica dell'Universit\`{a} di Milano
and INFN Sezione di Milano, \\
via Celoria , 16; 20133 Milano, Italy}


\maketitle\abstracts{In previous works,
a squared mass operator $ M^{2} $ and a more common
centre of frame Hamiltonian $ H_{\rm CM} $ for the
quark-antiquark system were obtained
by using the only assumption that
$ \ln W $ ($ W $ being the Wilson loop correlator) can be written
in QCD as the sum of its perturbative expression and an area term.
We evaluated the spectrum of these operators by neglecting all the
spin-dependent terms but for a spin-spin term. We succeeded in
reproducing rather well the entire meson spectrum and for the
light-light systems we found straight Regge trajectories with
the right slope and intercepts. A first attempt to study the
contribution of the spin dependent terms has been made.}

\section{Deduction of the quadratic and linear mass operators}

It was shown in previous works that starting from proper
two-particle and one-particle gauge invariant Green functions
$ H^{\rm gi}(x_{1},x_{2};y_{1},y_{2}) $ and
$ H^{\rm gi}(x-y) $ it is possible to obtain a first
principle Bethe-Salpeter and Dyson-Schwinger equation
respectively \cite{bmp}.
These gauge invariant Green functions are written in terms of a
``second order'' quark propagator
$ \Delta^{\sigma} (x,y;A) $ in a external gauge field
$ A^{\mu} $, defined as follow
by the iterated Dirac operator
\begin{equation}
 ( D_{\mu} D^{\mu} + m^{2} - \frac{1}{2} g \, \sigma^{\mu \nu}
 F_{\mu \nu} )
 \Delta^{\sigma} (x,y;A) = - \, \delta^{4} (x-y)
 \label{eq:propk}
 \end{equation}
with $ \sigma^{\mu \nu} = ( i/2 ) [ \gamma^{\mu}, \gamma^{\nu} ] $.
By a Feynmann-Schwinger representation,
$ \Delta^{\sigma} (x,y;A) $ can be written in terms of the Wilson
correlator
\begin{equation}
   W[\Gamma] = \frac{1}{3}
      \langle {\rm Tr} {\rm P} \exp \{ i g  \oint_\Gamma dx^{\mu}
      A_{\mu} \}  \rangle .
\end{equation}
In order to allow the evaluation of $ W $ we have to introduce
some assumptions. In the modified area law (MAL) model it was assumed
\begin{equation}
  i \ln W = i (\ln W)_{\rm pert} + \sigma S_{\rm min}
\label{eq:wl}
\end{equation}
where the first quantity is supposed to give correctly the short range
limit and the second the long range one.
If one replaces the minimal surface $ S_{\rm min} $ by its
``equal time straight line approximation'' and omits in the
path integral representation of
$ H^{\rm gi} ( x_{1},x_{2},y_{1},y_{2} ) $
and
$ H^{\rm gi} (x-y) $
the contributions to 
$ i \ln W $
coming from gluon lines and fermionic lines involving points
of the Schwinger strings, by an appropriate recurrence method,
an inhomogeneous Bethe-Salpeter equation and a Dyson-Schwinger
equation can be derived respectively.
The corresponding homogeneous BS-equation can be written as
\begin{eqnarray}
  \Phi_P (k) &=& - \, i \! \int \! \frac{ d^{4} u }{ (2 \pi)^{4} }
  \, \hat{I}_{ab} \big ( k - u, \frac{1}{2} P + \frac{ k + u }{2},
  \frac{1}{2} P - \frac{ k+u }{2} \big ) \nonumber \\
& & \qquad \quad \;
  \hat{H}_{1} \big( \frac{1}{2} P  + k \big) \, \sigma^{a} \,
  \Phi_{P} (u) \, \sigma^{b} \,
  \hat{H}_{2} \big( - \frac{1}{2} P + k \big)
\label{eq:bshoma}
\end{eqnarray}
and in terms of the irreducible self-energy
$ \hat{H} (k) = \hat{H}_{0} (k) + i \hat{H}_{0} (k) \hat{ \Gamma } (k)
 \hat{H} (k) $
the DS-equation can be written as
\begin{equation}
\hat{ \Gamma(k) } = \int \! \frac{ d^{4} l }{ (2 \pi)^{4} } \,
\hat{I}_{ab} \Big( k-l; \frac{k+l}{2}, \frac{k+l}{2} \Big) \,
\sigma^{a} \hat{H} (l) \, \sigma^{b} \, .
\label{eq:sdeq}
\end{equation}
The kernel $ \hat{I}_{ab} $ is completely known and because of
eq.(\ref{eq:wl}) it is given as the sum of a perturbative term and a
confinement term. In order to
solve eq.(\ref{eq:bshoma}) we should solve eq.(\ref{eq:sdeq})
and substitute $ \hat{H} $ into eq.(\ref{eq:bshoma}), then we
should solve eq.(\ref{eq:bshoma}). \\
We can simplify the problem by approximating  $ \hat{H} $ with a
free propagator $ -i/( p^{2} - m^{2} ) $ and performing an
instantaneous approximation that consists in setting
$ k_{0} = k_{0}^{\prime} = 
(1/2) [ \eta_{2} ( w_{1} + w_{1}^{\prime} ) -
\eta_{1} ( w_{2} + w_{2}^{\prime} ) ] $
with 
$ w_{j} = \sqrt{ m_{j}^{2} + {\bf k}^{2} } $,
$ w_{j}^{\prime} = \sqrt{ m_{j}^{2} + {\bf k}^{\prime 2} } $,
$ \eta_{j} = m_{j}/( m_{1} + m_{2} ) $
and
$ j = 1,2 $.
By performing explicitly the integration over $ k_{0}^{\prime} $
and $ k_{0} $ we obtain 
\begin{eqnarray}
  & &  ( w_{1} + w_{2} )^{2} \, \varphi_{ m_{B} } ( {\bf k} ) +
\label{autoval}
\\
  &+& \! \int \!
  \frac{ d^{3} {\bf k}^{\prime} }{ ( 2 \pi )^{3} }
  \, \sqrt{ \frac{ w_{1} + w_{2} }{ 2 w_{1} w_{2} } } \,
  \hat{I}_{ab}^{\rm inst} ( {\bf k}, {\bf k}^{\prime} )
  \, \sqrt{ \frac{ w_{1}^{\prime} + w_{2}^{\prime} }{ 2
  w_{1}^{\prime} w_{2}^{\prime} } } \, \sigma^{a} \,
  \varphi_{ m_{B} } ( {\bf k}^{\prime} ) \, \sigma^{b}
  = m_{B}^{2} \, \varphi_{ m_{B} } ( {\bf k} )
\nonumber
\end{eqnarray}
with
$
  \varphi_{P} ( {\bf k} ) =
  \sqrt{ 2 w_{1} w_{2} / ( w_{1} + w_{2} ) }
  \int_{ - \infty }^{\infty} \! d k_{0} \, \Phi_{P} ( { k } )
$.
Eq.(\ref{autoval}) is the eigenvalue equation  for the squared mass
operator
\begin{equation}
    M^2 = M_0^2 + U
\label{eq:quadr}
\end{equation}
with  $ M_0 = \sqrt{m_1^2 + {\bf k}^2} + \sqrt{m_2^2 + {\bf k}^2} \, $
and
\begin{equation}
  \langle {\bf k} | U | {\bf k}^{\prime} \rangle =
  \frac{1}{ (2 \pi)^3 }
  \sqrt{ \frac{ w_{1} + w_{2} }{ 2  w_{1}  w_{2} } }
  {\hat I}_{ab}^{\rm inst}
  ( {\bf k} , {\bf k}^{\prime} ) \sqrt{ 
  \frac{ w_{1}^{\prime} + w_{2}^{\prime} }{ 2
  w_{1}^{\prime} w_{2}^{\prime} } } \,
  \sigma_{1}^{a} \sigma_{2}^{b} \, .
\label{eq:quadrrel}
\end{equation}
The quadratic form of eq.(\ref{eq:quadr}) obviously derives from the second 
order character of the formalism we have used. 
In more usual terms one can also write
\begin{equation}
          H_{\rm CM} \equiv M = M_0 + V + \ldots 
\label{eq:lin}
\end{equation}
with
\begin{equation}
  \langle {\bf k} \vert V \vert {\bf k}^\prime \rangle =
  {1\over w_1 + w_2 + w_1^\prime + w_2^\prime }
  \langle {\bf k} \vert U \vert {\bf k}^\prime \rangle
  = \frac{1}{ (2 \pi)^3 }
  \frac{ \hat{I}_{ab}^{\rm inst} ( {\bf k}, {\bf k}^{\prime} ) }{ 4
  \sqrt{ w_{1} w_{2} w_{1}^{\prime} w_{2}^{\prime} } }
  \, \sigma_{1}^{a} \sigma_{2}^{b}
\label{eq:linrel}
\end{equation}
In eq.(\ref{eq:lin}) the dots stand for higher order terms in
$ \alpha_{\rm s} $
and $\sigma$; kinematic factors equal to 1 on the energy shell
have been neglected.

From eqs.(\ref{eq:quadrrel}) and (\ref{eq:quadr}) one obtains explicitly
\begin{eqnarray}
&&         \! \! \! \!
  \langle {\bf k} \vert U \vert {\bf k}^\prime \rangle =
  \sqrt{ \frac{ ( w_{1} +w_{2} ) ( w_{1}^{\prime}
  + w_{2}^{\prime} ) }{
  w_{1} w_{2} w_{1}^{\prime} w_{2}^{\prime} } }
  \bigg\{ {4\over 3} {\alpha_s \over \pi^2}
  \bigg[ - {1\over {\bf Q}^2}
  \bigg( q_{10} q_{20} + {\bf q}^2 -
  { ( {\bf Q} \cdot {\bf q})^2 \over {\bf Q}^2 } \bigg) +
\nonumber \\
&&  + \frac{i}{ 2 {\bf Q}^{2} }
  {\bf k} \times {\bf k}^{\prime} \cdot (
  \mbox{\boldmath $ \sigma $}_{1} +
  \mbox{\boldmath $ \sigma $}_{2} )
  + \frac{1}{ 2 {\bf Q}^2 } [ q_{20}
  ( \mbox{\boldmath $ \alpha $}_{1}
  \cdot {\bf Q} ) - q_{10} (
  \mbox{\boldmath $ \alpha $}_{2}
  \cdot {\bf Q}) ] +
\nonumber \\
&&  + \frac{1}{6}
  \mbox{\boldmath $ \sigma $}_{1}
  \cdot
  \mbox{\boldmath $ \sigma $}_{2}
  + \frac{1}{4} \left( \frac{1}{3}
  \mbox{\boldmath $ \sigma $}_{1}
  \cdot
  \mbox{\boldmath $ \sigma $}_{2} -
  \frac{ ( {\bf Q} \cdot
  \mbox{\boldmath $ \sigma $}_{1} )
  ( {\bf Q} \cdot
  \mbox{\boldmath $ \sigma $}_{2} ) }{
  {\bf Q}^2 } \right)
  + \frac{1}{ 4 {\bf Q}^2 } (
  \mbox{\boldmath $ \alpha $}_{1}
  \cdot {\bf Q} ) (
  \mbox{\boldmath $ \alpha $}_{2}
  \cdot {\bf Q} ) \bigg] +
\nonumber \\
&&  + \int \!
  \frac{ d^{3} {\bf r} }{ ( 2 \pi)^3 } \,
  e^{ i {\bf Q} \cdot {\bf r} }
  J^{\rm inst}( {\bf r} , {\bf q} , q_{10} , q_{20})  \bigg\}
\label{eq:upot}
\end{eqnarray}
with
\begin{eqnarray}
&& \! \! \! \! \! \!
  J^{\rm inst} ( {\bf r} , {\bf q} , q_{10} , q_{20} ) =
  \frac{ \sigma r }{ q_{10} + q_{20} }
  \bigg[ q_{20}^{2} \sqrt{ q_{10}^{2} - {\bf q}_{\rm T}^{2} }
  + q_{10}^{2} \sqrt{ q_{20}
  - {\bf q}_{\rm T}^{2} } +
\nonumber  \\
&& \! \! \! \! \! \!
  + \frac{ q_{10}^{2} q_{20}^{2} }{ \vert {\bf q}_{\rm T} \vert }
  \big( \arcsin \frac{ \vert {\bf q}_{\rm T} \vert }{ q_{10} }
  + \arcsin \frac{ \vert {\bf q}_{\rm T} \vert }{ q_{20} }
  \bigg) \bigg]
  - \frac{ \sigma }{r} \bigg[ \frac{ q_{20} }{
  \sqrt{ q_{10}^{2} - {\bf q}^{2}_{\rm T} } }
  ( {\bf r} \times {\bf q} \cdot
  \mbox{\boldmath $ \sigma $}_{1} +
\nonumber  \\
&& \! \! \! \! \! \!
  + i q_{10} ( {\bf r} \cdot
  \mbox{\boldmath $ \alpha $}_{1} ) )
  + \frac{ q_{10} }{ \sqrt{ q_{20}^{2} - {\bf q}^{2}_{\rm T} } }
  ( {\bf r} \times {\bf q} \cdot
  \mbox{\boldmath $ \sigma $}_{2}
  - i q_{20} ( {\bf r} \cdot
  \mbox{\boldmath $ \alpha $}_{2} ) ) \bigg]
\label{eq:uconf1}
\end{eqnarray}
here
$
  \alpha_{j}^{k} = \gamma_{j}^{0} \gamma_{j}^{k}, \,
  \sigma_{j}^{k} = \frac{i}{4} \,
  \varepsilon^{knm} [ \gamma^{n}_{j} , \gamma^{m}_{j} ], \,
  {\bf Q} = {\bf k} - {\bf k}^{\prime}, \,
  {\bf q} = ( {\bf k} + {\bf k}^{\prime} )/2
$
and
$ q_{j0} = ( w_{j} + w_{j}^{\prime} )/2 $. \\

Due to eq.(\ref{eq:linrel}) the potential $ V $ can be obtained from
$ U $ as given by eqs.(\ref{eq:upot})-(\ref{eq:uconf1})
simply by the kinematic replacement
$
  \sqrt{ \frac{ ( w_{1} + w_{2} )
  ( w_{1}^{\prime} + w_{2}^{\prime} ) }{ w_{1} w_{2}
  w_{1}^{\prime} w_{2}^{\prime} } }
\to \frac{1}{ 2\sqrt{ w_{1} w_{2} w_{1}^{\prime} w_{2}^{\prime} } }
$.

\section{Quarkonium spectrum and Regge trajectories}

We computed the heavy-heavy, heavy-light and light-light
quarkonium spectrum and the the Regge trajectories for the
light-light quarkonium systems by
using both the linear mass operator $ H_{\rm CM} $
and the quadratic mass operator $ M^{2} $.
We  solved their eigenvalue
equation by the Rayleigh-Ritz variational method
using the three-dimensional harmonic oscillator basis
\cite{simon}.
In the linear potential $ V $ and in the quadratic interaction
$ U $ we have taken into account only the velocity dependent terms
in order to compute the baricentral meson masses and the spin-spin
$ \sigma_{1} \cdot \sigma_{2} $ term in order to compute the
hyperfine splitting. We have then neglected all the other spin
dependent terms that will be studied later.
\begin{figure}
  \setlength{\unitlength}{1.0mm}
  \begin{picture}(100,100)(5,0)
    \put(0,60){\mbox{\epsfig{file=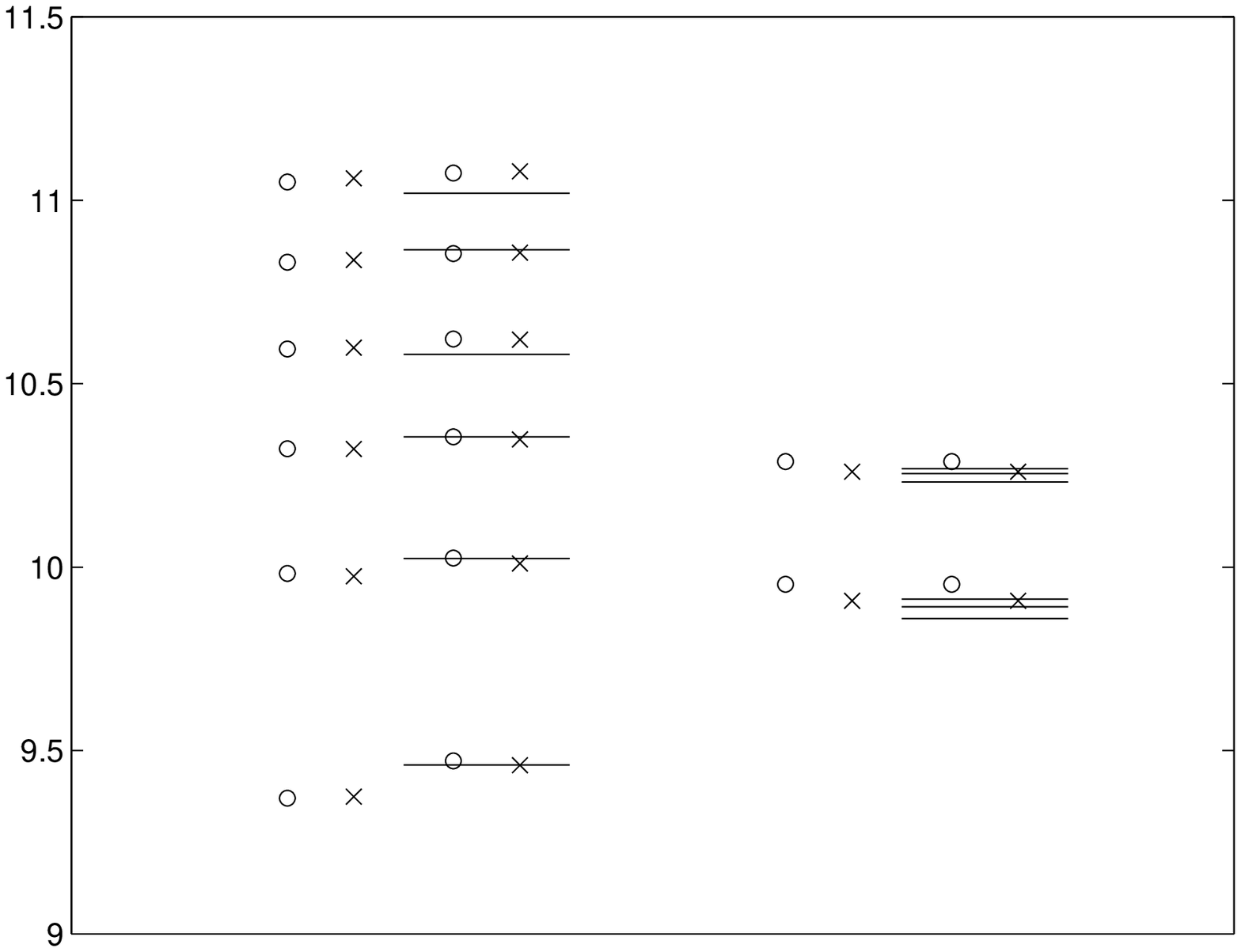,height=5cm,width=6cm}}}
    \put(70,60){\mbox{\epsfig{file=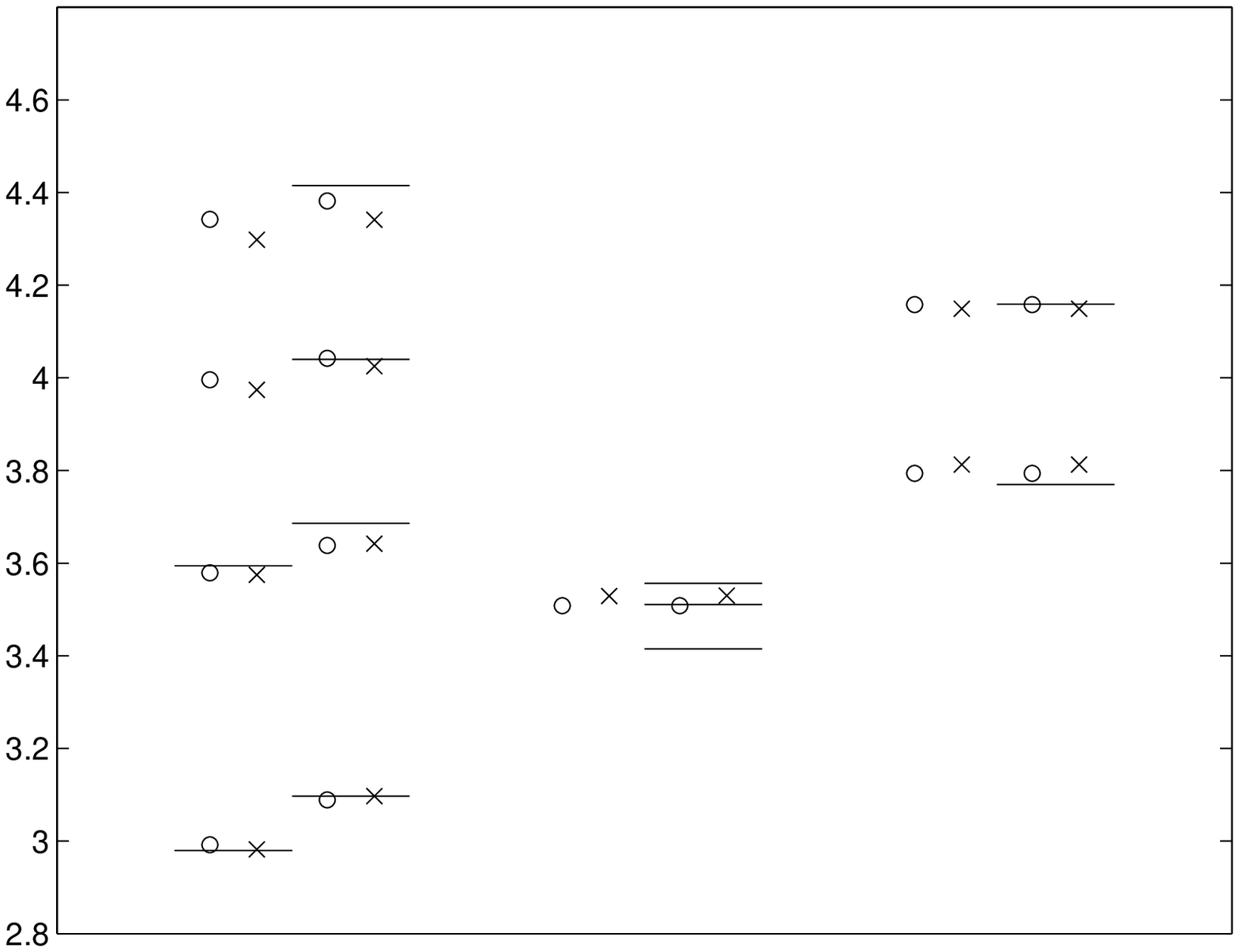,height=5cm,width=6cm}}}
    \put(0,0){\mbox{\epsfig{file=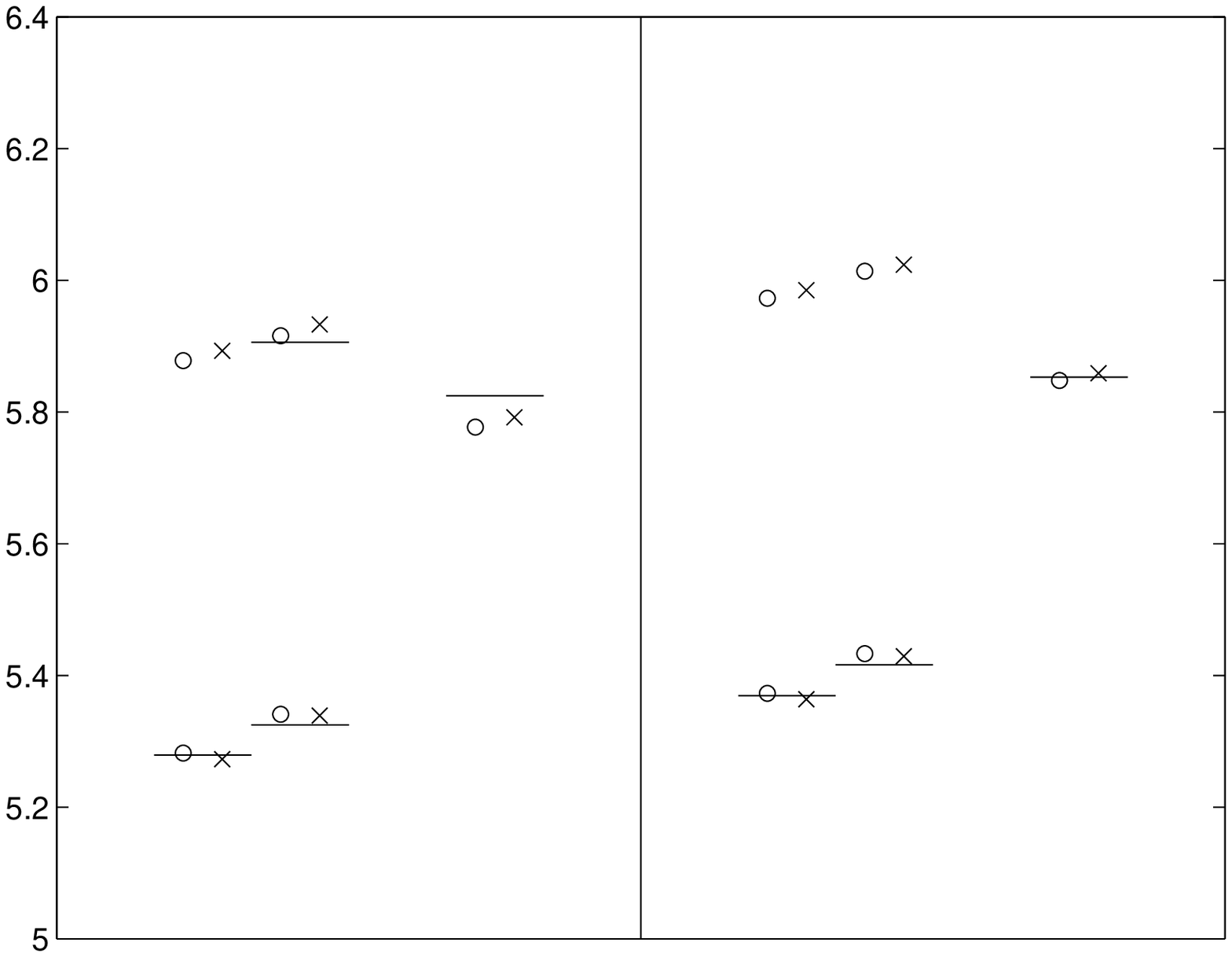,height=5cm,width=6cm}}}
    \put(70,0){\mbox{\epsfig{file=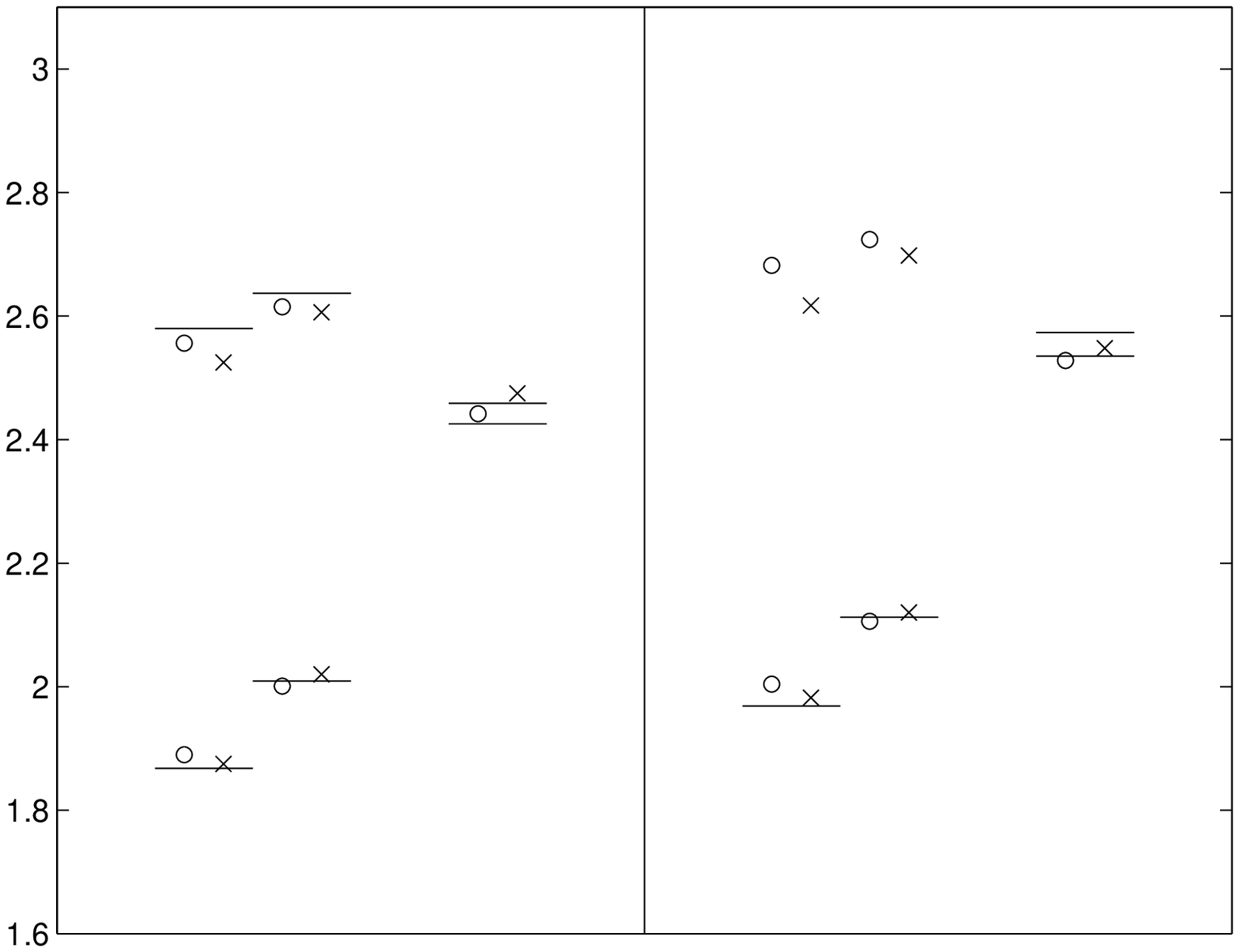,height=5cm,width=6cm}}}
    \put(46,68){ $ b \bar{b} $ }
    \put(12,105){ $ {}_{^{1} {\rm S}_{0}} $ }
    \put(20,105){ $ {}_{^{3} {\rm S}_{1}} $ }
    \put(28,105){ $ {}_{n} $ }
    \put(28,100){ $ {}_{{}_{6}} $ }
    \put(28,97){ $ {}_{{}_{5}} $ }
    \put(28,92){ $ {}_{{}_{4}} $ }
    \put(28,87){ $ {}_{{}_{3}} $ }
    \put(28,80.5){ $ {}_{{}_{2}} $ }
    \put(28,70){ $ {}_{{}_{1}} $ }
    \put(36,105){ $ {}_{^{1} {\rm P}_{1}} $ }
    \put(44,105){ $ {}_{^{3} {\rm P}_{J}} $ }
    \put(52,105){ $ {}_{n} $ }
    \put(52,85){ $ {}_{{}_{2}} $ }
    \put(52,78){ $ {}_{{}_{1}} $ }
    \put(115,68){ $ c \bar{c} $ }
    \put(77,105){ $ {}_{^{1} {\rm S}_{0}} $ }
    \put(84,105){ $ {}_{^{3} {\rm S}_{1}} $ }
    \put(90,105){ $ {}_{n} $ }
    \put(90,100){ $ {}_{{}_{4}} $ }
    \put(90,91){ $ {}_{{}_{3}} $ }
    \put(90,81){ $ {}_{{}_{2}} $ }
    \put(90,67){ $ {}_{{}_{1}} $ }
    \put(94,105){ $ {}_{^{1} {\rm P}_{1}} $ }
    \put(101,105){ $ {}_{^{3} {\rm P}_{J}} $ }
    \put(107,105){ $ {}_{n} $ }
    \put(107,78.5){ $ {}_{{}_{1}} $ }
    \put(111,105){ $ {}_{^{1} {\rm D}_{1}} $ }
    \put(118,105){ $ {}_{^{3} {\rm D}_{J}} $ }
    \put(125,100){ $ {}_{{}_{J}} $ }
    \put(125,94){ $ {}_{{}_{1}} $ }
    \put(125,85){ $ {}_{{}_{1}} $ }
    \put(22,7){ $ u \bar{b} $ }
    \put(6,45){ $ {}_{^{1} {\rm S}_{0}} $ }
    \put(12,45){ $ {}_{^{3} {\rm S}_{1}} $ }
    \put(20,45){ $ {}_{^{3} {\rm P}_{J}} $ }
    \put(51,7){ $ s \bar{b} $ }
    \put(35,45){ $ {}_{^{1} {\rm S}_{0}} $ }
    \put(41,45){ $ {}_{^{3} {\rm S}_{1}} $ }
    \put(49,45){ $ {}_{^{3} {\rm P}_{J}} $ }
    \put(92,7){ $ u \bar{c} $ }
    \put(76,45){ $ {}_{^{1} {\rm S}_{0}} $ }
    \put(82,45){ $ {}_{^{3} {\rm S}_{1}} $ }
    \put(90,45){ $ {}_{^{3} {\rm P}_{J}} $ }
    \put(97,40){ $ {}_{{}_{J}} $ }
    \put(97,29.5){ $ {}_{{}_{2}} $ }
    \put(97,27.5){ $ {}_{{}_{1}} $ }
    \put(121,7){ $ s \bar{c} $ }
    \put(105,45){ $ {}_{^{1} {\rm S}_{0}} $ }
    \put(111,45){ $ {}_{^{3} {\rm S}_{1}} $ }
    \put(119,45){ $ {}_{^{3} {\rm P}_{J}} $ }
    \put(125.5,40){ $ {}_{{}_{J}} $ }
    \put(125.5,33){ $ {}_{{}_{2}} $ }
    \put(125.5,31){ $ {}_{{}_{1}} $ }
  \end{picture}
\caption{Heavy-heavy and heavy-light quarkonium spectrum.
Circlets are our results by the $H_{\rm CM}$ operator,
crosses are our results by the $M^{2}$ operator.}
\label{fig1}
\end{figure}
\begin{figure}
  \setlength{\unitlength}{1.0mm}
  \begin{picture}(100,150)(5,0)
    \put(0,120){\mbox{\epsfig{file=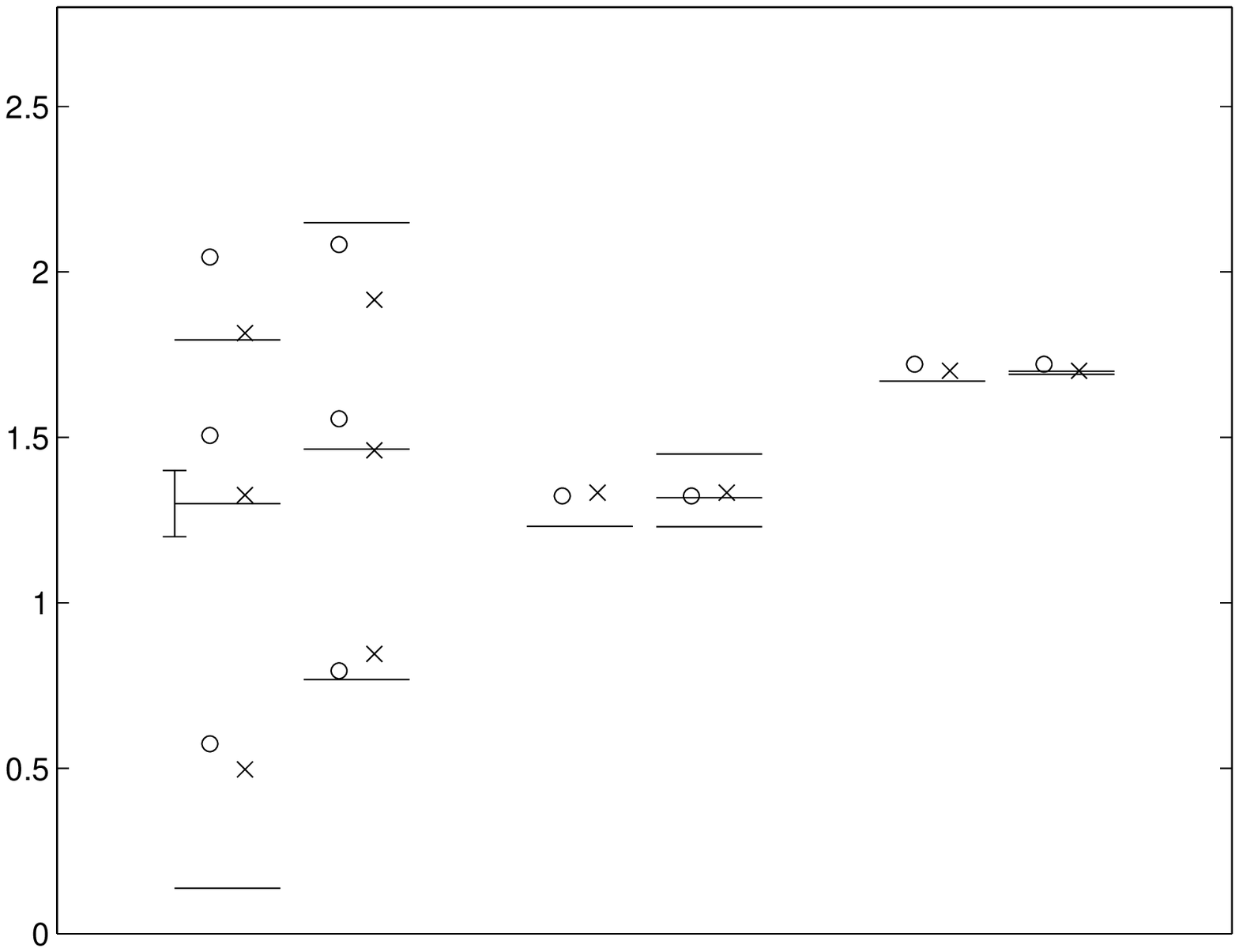,height=5cm,width=6cm}}}
    \put(70,120){\mbox{\epsfig{file=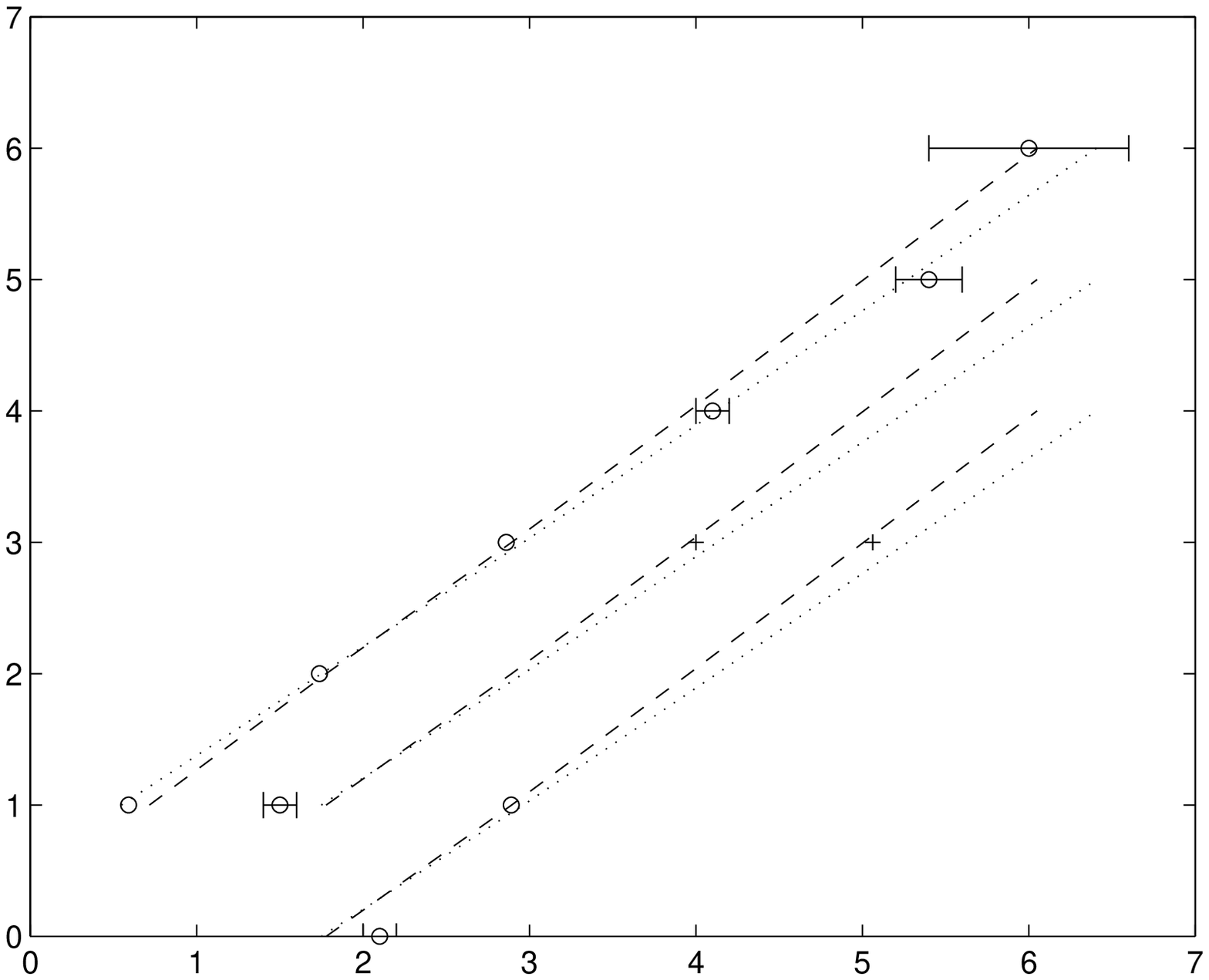,height=5cm,width=6cm}}}
    \put(0,60){\mbox{\epsfig{file=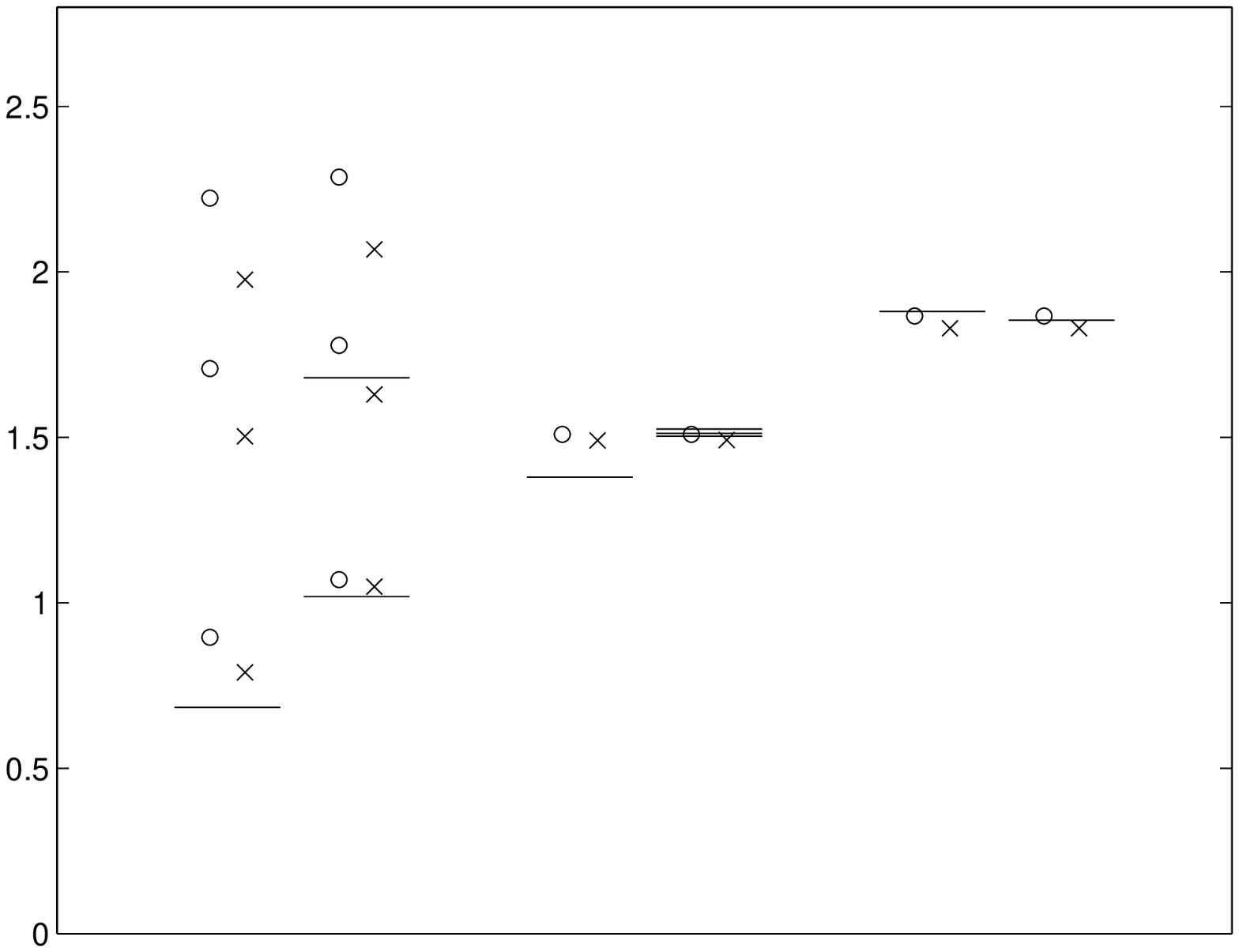,height=5cm,width=6cm}}}
    \put(70,60){\mbox{\epsfig{file=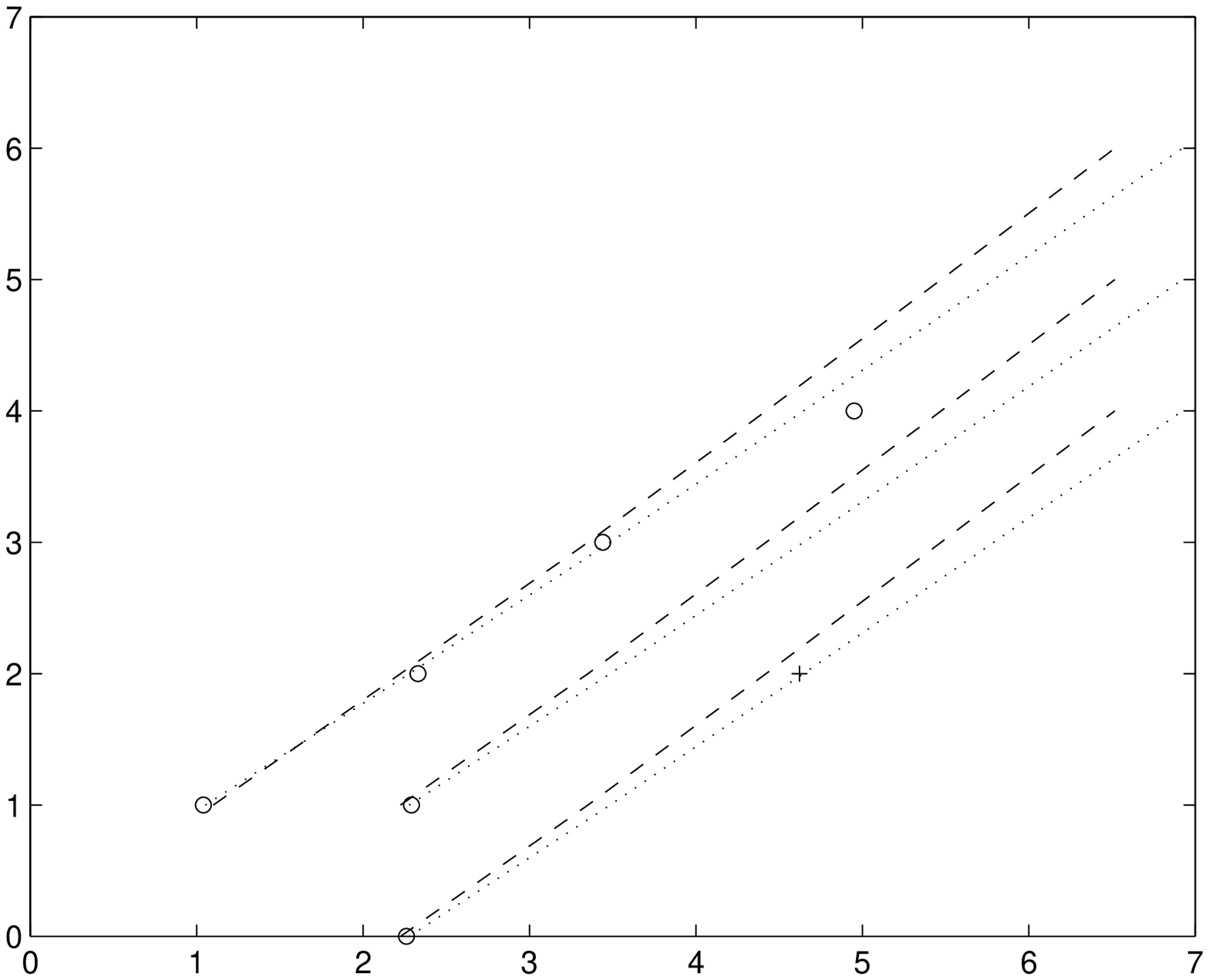,height=5cm,width=6cm}}}
    \put(0,0){\mbox{\epsfig{file=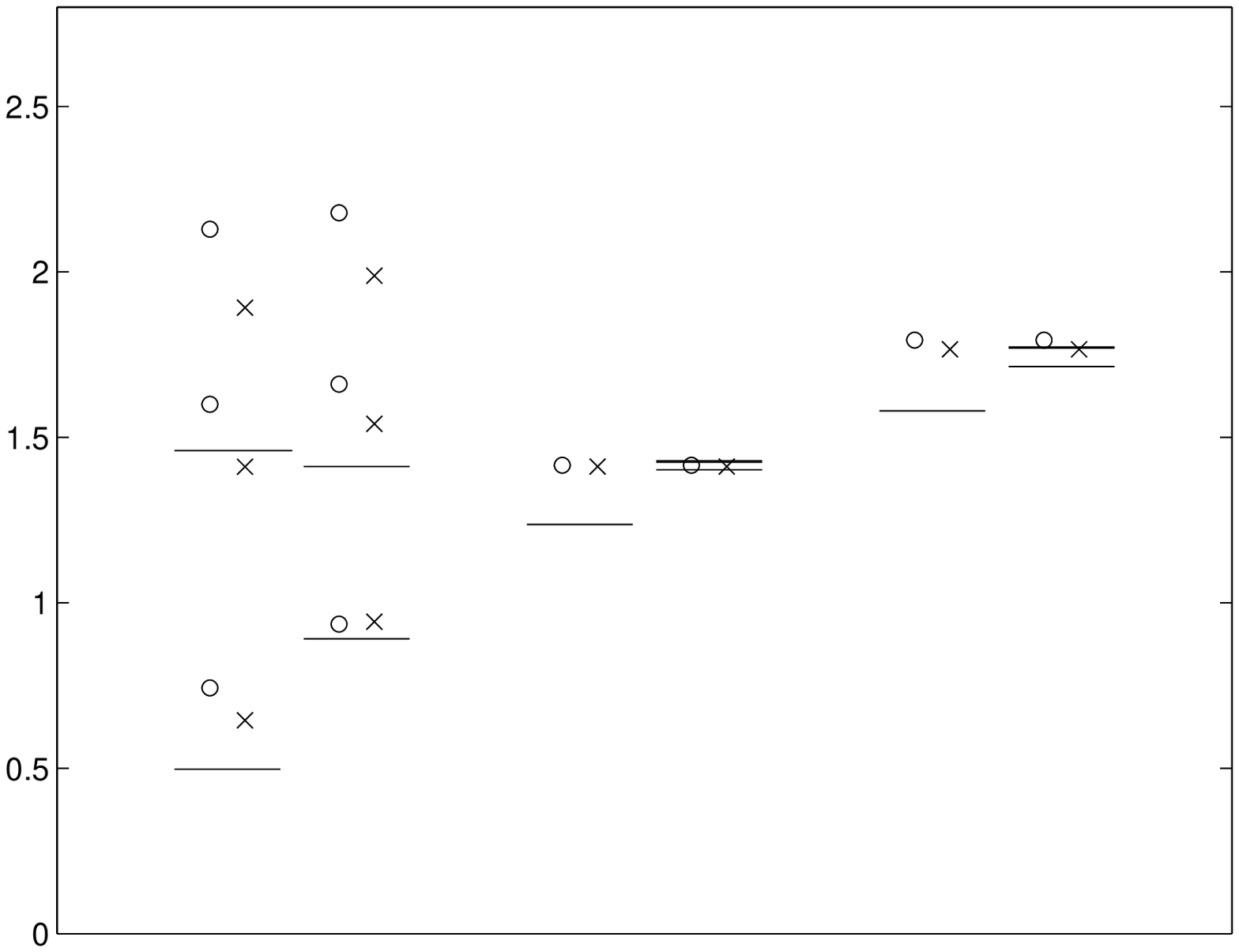,height=5cm,width=6cm}}}
    \put(70,0){\mbox{\epsfig{file=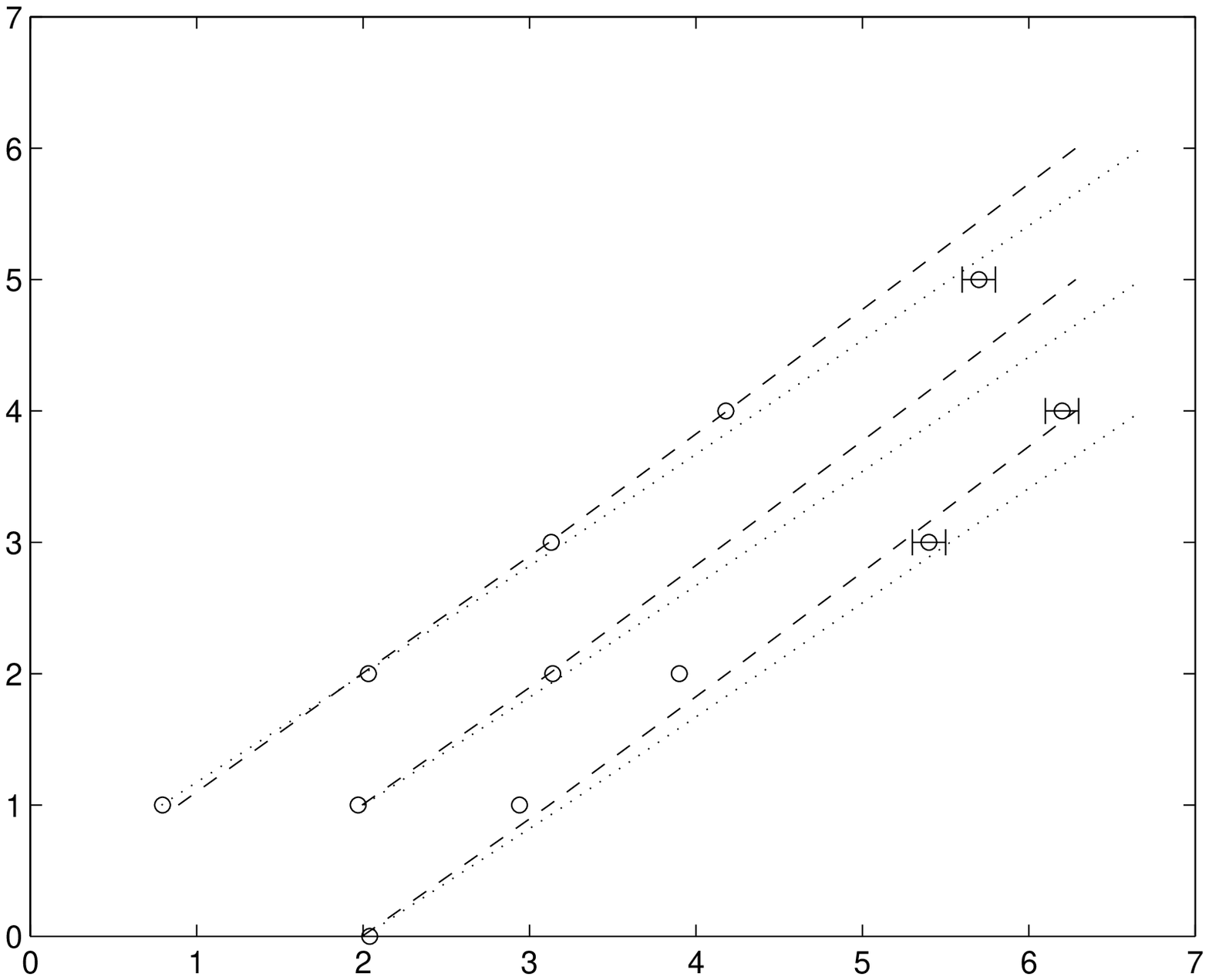,height=5cm,width=6cm}}}
    \put(48,128){ $ u \bar{u} $ }
    \put(7,165){ $ {}_{^{1} {\rm S}_{0}} $ }
    \put(14,165){ $ {}_{^{3} {\rm S}_{1}} $ }
    \put(24,165){ $ {}_{^{1} {\rm P}_{1}} $ }
    \put(31,165){ $ {}_{^{3} {\rm P}_{J}} $ }
    \put(42,165){ $ {}_{^{1} {\rm D}_{1}} $ }
    \put(49,165){ $ {}_{^{3} {\rm D}_{J}} $ }
    \put(55,159){ $ {}_{{}_{J}} $ }
    \put(54,151){ $ {}_{{}_{1,3}} $ }
    \put(80,162){ $ u \bar{u} $ }
    \put(67.5,165){ $ {}_{J} $ }
    \put(129,121){ $ {}_{M^{2}} $ }
    \put(124,164){ $ {}_{J = L + 1} $ }
    \put(124,157){ $ {}_{J = L} $ }
    \put(124,150){ $ {}_{J = L - 1} $ }
    \put(68,131){ $ {}_{\rho ( 770 )} $ }
    \put(73,137){ $ {}_{a_{2} ( 1320 )} $ }
    \put(82,143){ $ {}_{\rho_{3} ( 1690 )} $ }
    \put(91,150){ $ {}_{a_{4} ( 2040 )} $ }
    \put(101,157){ $ {}_{\rho_{5} ( 2350 )} $ }
    \put(108,164){ $ {}_{a_{6} ( 2450 )} $ }
    \put(74,126){ $ {}_{a_{1} ( 1260 )} $ }
    \put(96,140){ $ {}_{X( 2000 )} $ }
    \put(83,119){ $ {}_{a_{0} ( 1450 )} $ }
    \put(95,128){ $ {}_{\rho ( 1700 )} $ }
    \put(113,141){ $ {}_{\rho_{3} ( 2250 )} $ }
    \put(48,68){ $ s \bar{s} $ }
    \put(7,105){ $ {}_{^{1} {\rm S}_{0}} $ }
    \put(14,105){ $ {}_{^{3} {\rm S}_{1}} $ }
    \put(24,105){ $ {}_{^{1} {\rm P}_{1}} $ }
    \put(31,105){ $ {}_{^{3} {\rm P}_{J}} $ }
    \put(42,105){ $ {}_{^{1} {\rm D}_{1}} $ }
    \put(49,105){ $ {}_{^{3} {\rm D}_{J}} $ }
    \put(55,99){ $ {}_{{}_{J}} $ }
    \put(55,94){ $ {}_{{}_{3}} $ }
    \put(14,73){ $ {}_{\eta_{S}} $ }
    \put(80,102){ $ s \bar{s} $ }
    \put(67.5,105){ $ {}_{J} $ }
    \put(129,61){ $ {}_{M^{2}} $ }
    \put(124,104){ $ {}_{J = L + 1} $ }
    \put(124,97){ $ {}_{J = L} $ }
    \put(124,90){ $ {}_{J = L - 1} $ }
    \put(70.5,71){ $ {}_{\phi ( 1020 )} $ }
    \put(77.5,77){ $ {}_{f_{2}^{\prime} ( 1525 )} $ }
    \put(87,84){ $ {}_{\phi_{3} ( 1850 )} $ }
    \put(97,91){ $ {}_{f_{j} ( 2220 )} $ }
    \put(82,67){ $ {}_{f_{1} ( 1510 )} $ }
    \put(86,58.7){ $ {}_{f_{0} ( 1500 )} $ }
    \put(109,74){ $ {}_{f_{2} ( 2150 )} $ }
      \put(48,8){ $ u \bar{s} $ }
      \put(7,45){ $ {}_{^{1} {\rm S}_{0}} $ }
      \put(14,45){ $ {}_{^{3} {\rm S}_{1}} $ }
      \put(24,45){ $ {}_{^{1} {\rm P}_{1}} $ }
      \put(31,45){ $ {}_{^{3} {\rm P}_{J}} $ }
      \put(42,45){ $ {}_{^{1} {\rm D}_{1}} $ }
      \put(49,45){ $ {}_{^{3} {\rm D}_{J}} $ }
      \put(55,39){ $ {}_{{}_{J}} $ }
      \put(54,33){ $ {}_{{}_{2,3}} $ }
      \put(55,30){ $ {}_{{}_{1}} $ }
    \put(80,42){ $ u \bar{s} $ }
    \put(67.5,45){ $ {}_{J} $ }
    \put(129,1){ $ {}_{M^{2}} $ }
    \put(124,44){ $ {}_{J = L + 1} $ }
    \put(124,37){ $ {}_{J = L} $ }
    \put(124,30){ $ {}_{J = L - 1} $ }
    \put(68,11.5){ $ {}_{K^{\ast} ( 892 )} $ }
    \put(74.5,17){ $ {}_{K_{2}^{\ast} ( 1430 )} $ }
    \put(83,23){ $ {}_{K_{3}^{\ast} ( 1780 )} $ }
    \put(92,30){ $ {}_{K_{4}^{\ast} ( 2045 )} $ }
    \put(102,37){ $ {}_{K_{5}^{\ast} ( 2380 )} $ }
    \put(78,6.5){ $ {}_{K_{1} ( 1400 )} $ }
    \put(88,13){ $ {}_{K_{2} ( 1770 )} $ }
    \put(108,22){ $ {}_{\longleftarrow} $ }
    \put(116,21){ $ {}_{K_{3} ( 2320 )} $ }
    \put(117,28.5){ $ {}_{\leftarrow} $ }
    \put(116,26){ $ {}_{K_{4} ( 2500 )} $ }
    \put(82,-1.5){ $ {}_{K_{0}^{\ast} ( 1430 )} $ }
    \put(96,7){ $ {}_{K^{\ast} ( 1680 )} $ }
    \put(102.5,15.5){ $ {}_{\rightarrow} $ }
    \put(106,14){ $ {}_{K_{2}^{\ast} ( 1980 )} $ }
  \end{picture}
\caption{Light-light quarkomium spectrum and Regge trajectories.
Circlets and dotted lines are our results by the $H_{\rm CM}$ operator,
crosses and dashed lines are our results by the $M^{2}$ operator.}
\label{fig2}
\end{figure}
As shown in figs.\ref{fig1}-\ref{fig2} our theoretical meson masses
are in good agreement with all the experimental data but for the light
pseudo-scalar states which are related to the chiral symmetry problematic. 
We found also straight Regge trajectories with the right slope and
intercepts. \\
The introduction of a running coupling constant was especially
required in the computation of the $ M^{2} $ eigenvalues. However it
was introduced also in the $ H_{\rm CM} $ case. We chose the simplest
perturbative running coupling constant
$
  \alpha_{\rm s}(\bf Q) =
  ( 4 \pi )/[ (11 - \frac{ 2 }{ 3 } N_{\rm f} )
  \ln { ( {\bf Q}^{2} / \Lambda^{2} ) }  ],
$
with $ N_{\rm f} = 4, \;  \Lambda = 200 \, {\rm MeV} $ 
and a horizontal cut at a value
$ \alpha_{\rm s}(0) $
in order to avoid the singular point
$ {\bf Q}^{2} = \Lambda^{2} $.
We chose: the heavy quark masses $ m_{\rm b} $ and
$ m_{\rm c} $, in order to reproduce exactly the masses of
$ J/\Psi $ and $ \Upsilon $; 
the running coupling constant cut $ \alpha_{\rm s}(0) $,
in order to reproduce the charmonium hyperfine splitting;
the string tension $ \sigma $,
in order to obtain a right slope for the
$ u \bar{u} $ Regge trajectories.
No fit was made on the light-heavy quarkonium systems.
\\
The light quark masses were fixed on typical current values
\cite{prtdatb}
$ m_{\rm u} = m_{\rm d} = 10 \, {\rm MeV} $ and
$ m_{\rm s} = 200 \, {\rm MeV} $
both for the first order formalism and for the second order
formalism.
\\
In the first order formalism we adopted the following
parameters \cite{lineare,quadratico}:
$ m_{\rm c} = 1.40 \, {\rm GeV} $,
$ m_{\rm b} = 4.81 \, {\rm GeV} $,
$ \alpha_{\rm s}(0) = 0.363 $,
$ \sigma = 0.175 \, {\rm GeV}^{2} $; whereas in the
second order formalism we adopted the following parameters
\cite{quadratico}:
$ m_{\rm c} = 1.394 \, {\rm GeV} $,
$ m_{\rm b} = 4.763 \, {\rm GeV} $,
$ \alpha_{\rm s}(0) = 0.35 $,
$ \sigma = 0.2 \, {\rm GeV}^{2} $.

A first attempt to study the contribution of the spin dependent
terms has been made.
For the  heavy-light systems a lot of spin dependent terms
are negligible and we are left with a small number of them.
The problem becomes very similar to the study of the hydrogen-like
atom
spectrum by means of the quadratic form of the Dirac equation
\cite{dirac}.
This part of the work is in progress.


\section*{References}
\begin{thebibliography}{99}
\bibitem{bmp}
  N. Brambilla, E. Montaldi and G. M. Prosperi, {\sl Phys. Rev.}
  {\bf D54} (1996) 3506; \hspace{2mm}
  G.M. Prosperi, hep-th/9709046;
  proc. XI int. conf.
  {\sl Problems in Quantum Theory of Fields}, Dubna 1998.
\bibitem{simon}
  M. Reed and B. Simon, {\sl Methods of Modern Mathematical Physics
  IV: Analysis of Operators} (Academic Press, New York, 1978) Section
  XIII.1 and XIII.2; \hspace{2mm}
  E.J. Weniger, {\sl J. Math. Phys.} {\bf 26} (2) (1985) 276.
\bibitem{prtdatb}
  Particle Data Group, R.M. Barnett {\sl et al.},
  Phys. Rev. D 54 (1996).
\bibitem{lineare}
  M. Baldicchi and G.M. Prosperi, {\sl Phys. Lett.}
  {\bf B 436} (1998), 145; \\
  M. Baldicchi and G.M. Prosperi, proc.
  {\sl Quark Confinement and the Hadron Spectrum III},
  World Scientific, Singapore (in print).
\bibitem{quadratico}
  M. Baldicchi and G.M. Prosperi, {\sl Fizika}
  {\bf B 8} (1999) 2, 251.
\bibitem{dirac}
  C. Itzykson and J.B. Zuber, Quantum Field Theory,
  Sects. 2-3-1, 2-3-2.
\end{thebibliography}
\end{document}